\documentclass{aastex}
\usepackage{emulateapj5}

\newcommand{\sauron}{{\texttt {SAURON}}}

\begin{document}

\title{Stellar kinematics in double-barred galaxies:
 the \emph{$\sigma$}-hollows}
\author{A.~de Lorenzo-C\'aceres\altaffilmark{1},
J.~Falc\'on-Barroso\altaffilmark{2}, 
A.~Vazdekis\altaffilmark{1} and 
I.~Mart\'inez-Valpuesta\altaffilmark{1}}
\altaffiltext{1}{Instituto de Astrof\'isica de Canarias (IAC)}
\altaffiltext{2}{European Space Agency (ESA)/ESTEC}

\begin{abstract}
We present \sauron\ integral-field stellar velocity and velocity dispersion
maps for four double-barred early-type galaxies: NGC\,2859, NGC\,3941, NGC\,4725
and NGC\,5850. The presence of the inner bar does not produce major changes in the line-of-sight 
velocity, but it appears to have an important effect in the stellar velocity 
dispersion maps: we find two \emph{$\sigma$}-hollows of amplitudes 
between 10 and 40 km~s$^{-1}$ on either side of the center, at the ends of the inner 
bars. We have performed numerical simulations to explain these features. 
Ruling out other possibilities, we conclude that the \emph{$\sigma$}-hollows are an
effect of the contrast between two kinematically different components: the high velocity 
dispersion of the bulge and the more ordered motion (low velocity dispersion) of the inner bar.
\end{abstract}


\section{Introduction}
Double-barred galaxies are rather common systems in the Universe, 
representing $\sim$1/3 of all barred galaxies
\citep{2002AJ....124...65E,2002ApJ...567...97L}. These structures are mainly
seen by surface brightness profile decomposition, which requires not
only a bulge and a disk but also two additional components with non-axial
symmetry. The presence of inner bars does not seem to correlate with the 
morphological type of the host galaxy and there is also no preferred
angle between the two structures \citep{1993A&A...277...27F,1995A&AS..111..115W}. 
This property suggests that both bars rotate independently, as seen in 
observations and consistent with numerical simulations 
\citep[e.g.][]{2003ApJ...599L..29C,2004ApJ...617L.115E,1989Natur.338...45S,2002ApJ...565..921S}. 
Inner bars play an important role in the evolution of galaxies as they might 
transport gas to the central regions, where they may trigger the 
formation of stars that, in turn, lead to the appearance of new structures 
such as disk-like components \citep{1995FCPh...15..341M}. Theoretically, this 
phenomenon cannot be achieved with a single bar, since the flow of material 
would stop before reaching the galactic center (e.g. at the Inner Lindblad resonance). 
The presence of two embedded bars makes it possible and may even help to 
feed the central AGN \citep{1989Natur.338...45S,1990Natur.345..679S}. 

The study of the orbital make-up of different morphological components in 
double-barred galaxies is one of the best ways to understand their internal 
structure and dynamics. Nevertheless, this is a rather complex issue to address as 
there are three fundamental frequencies for the regular orbits \citep[i.e. two 
for the bars and one for the free oscillations;][]{MA07} instead of the two 
frequencies as in the case of a single bar. As a result, they do not have closed 
and periodic orbits, so \citet{2000MNRAS.313..745M} introduced the 
\emph{loop} concept to represent the stable orbits in a double-barred potential.

A more popular approach is the use of numerical 
simulations to describe the structures seen in real observations, such as
in the recent works by \citet{2007ApJ...657L..65H} and \citet{2007ApJ...654L.127D}.
In their simulations, these authors generate long-lived double bars 
with and without a dissipative component, 
respectively. However, most existing observations have relied on long-slit spectroscopy along 
only a few position angles \citep[e.g.][]{2001A&A...368...52E}, which makes the comparison with 
the models somewhat difficult.
Better suited integral-field spectroscopy has generally not been used to observe these systems,
with the notable exceptions of \citet{2001BSAO...51..140M} and \citet{2004A&A...421..433M}.

Here we present \sauron\ integral-field spectroscopy results on the stellar 
kinematics of four double-barred early-type galaxies. We will 
focus on the detailed analysis of the stellar velocity dispersion maps, that 
reveal a peculiar feature, not seen before, along the extremes of the inner 
bar. We present the observational evidence and discuss possible interpretations 
using numerical simulations. A more complete account of the ionised gas,
stellar population properties and supporting numerical
simulations will be the subject of a forthcoming paper.

\section{Observations, instrumental setup and analysis}
Integral-field spectroscopic data of four double-barred early-type galaxies were
taken with the \sauron\ spectrograph, attached to the William Herschel Telescope
at the Observatorio del Roque de los Muchachos (La Palma, Spain). The galaxies
were selected from the catalog of \citet{2004A&A...415..941E}, so that not 
only the lengths of the inner bars were a good match to the \sauron\ field-of-view (FoV), 
but also allowed for enough FoV in order to sample the transition region between the two bars. 
We deliberately selected early-type galaxies so as to avoid the appearance of complex 
structures due to the presence of dust, and to make it easier to perform 
stellar population analysis.

We used the LR mode of \sauron\, providing a $33\arcsec\times41\arcsec$ 
FoV with spatial sampling of $0\farcs94\times0\farcs94$ per lens. The spectral 
range covers the domain between 4800-5380~\AA. This setup produces 1431 spectra 
per pointing over the \sauron\ FoV, with a sampling of 1.1 \AA\,pixel$^{-1}$ 
and a spectral resolution of 3.74~\AA\,(FWHM). The data reduction was performed using 
the {\tt XSAURON} package \citep{2001MNRAS.326...23B}, following the same
procedures outlined in \citet{2004MNRAS.352..721E}. Total integration times 
range between 3 to 4 hours, providing a signal-to-noise ratio of $\sim300$ at the 
centers of the galaxies. Additionally, in order to ensure the measurement of 
reliable stellar kinematics, we spatially binned our final datacubes using the 
Voronoi 2D binning algorithm of \citet{2003MNRAS.342..345C}, creating compact 
bins with a minimum signal-to-noise ratio ($S/N$) of $\sim60$ per spectral 
resolution element. Most spectra in the central regions, however, have $S/N$ in 
excess of $60$, and so remain un-binned. The stellar kinematics is then extracted as described in
\citet{2006MNRAS.369..529F}, by fitting the absorption spectrum to a linear
combination of the same set of templates. For this purpose we used the penalized 
pixel-fitting (pPXF) method of \citet{2004PASP..116..138C} and a well selected 
subsample of the stellar population models from \citet{1999ApJ...513..224V},
with evenly sampled ages and metallicities.

\section{The observed $\sigma$-hollows}
\begin{figure}[ttt]
\begin{center}
\includegraphics[angle=180,width=12cm]{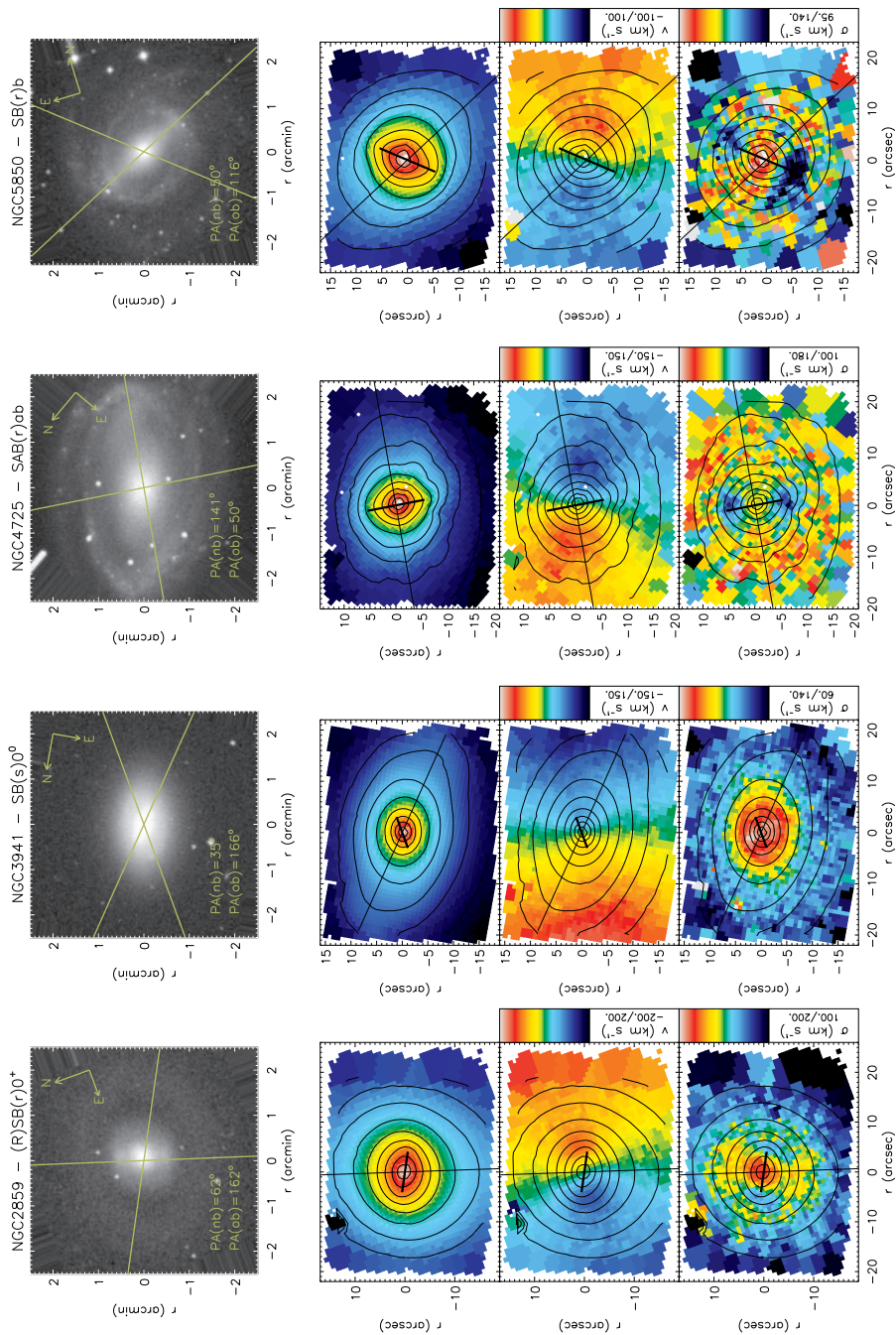}
\caption{Maps of the stellar distribution and kinematics of the 4 double-barred 
galaxies in our sample: an optical image from the Sloan Digital Sky Survey 
\citep{2000AJ....120.1579Y} together with our intensity maps and the stellar 
velocity and stellar velocity dispersion maps are shown for each galaxy. 
We have overplotted the position angle of the 
inner bar (thick line), the outer bar (thin line) and the contours of the
reconstructed total intensity from our own datacubes. Note the presence of 
the $\sigma$-hollows at the edges of the four inner bars; these hollows are clearly seen in
the velocity dispersion maps of NGC\,2859, NGC\,4725 and NGC\,5850 whereas they are less evident 
(that is, they have lower amplitudes and their presence is evident from cuts along the
inner bar) for the case of NGC\,3941.}
\label{fig:data}
\end{center}
\end{figure}

Figure~\ref{fig:data} shows the results from the analysis of the datacubes. The 
stellar velocity maps look very much like those of non double-barred galaxies and in
three out of the four galaxies (NGC\,2859, NGC\,4725 and NGC\,5850) we find 
local velocity maxima and minima along the direction of the kinematic major axes, at a few
arcseconds from the galaxy centers. Moreover, the $h_3$ maps, not included in this letter, show a 
clear anticorrelation with the stellar velocities at the locations of these features, 
supporting the idea that these faster rotating components are kinematically decoupled
inner disks \citep{2005ApJ...626..159B}. Since the kinematic major axes 
are not aligned with the inner bars at these regions, these disks cannot be related with 
the bars.

However, the stellar velocity dispersion maps reveal some interesting features
not seen in other morphological types of galaxies. The values measured are at a 
maximum at the centers of the galaxies (as seen in most E/S0 galaxies) but, 
instead of smooth negative velocity dispersion gradients towards the outer 
parts, we find two symmetrical regions where the velocity dispersion values drop 
significantly compared to their surroundings. These $\sigma$-hollows are located
exactly along the major axes of the inner bars and they extend out to the edges,
as we checked with several tests. The amplitude of the hollows varies between
10 and 40 km~s$^{-1}$. Interestingly, neither the H$_{\beta}$ or $[$OIII$]$
emission-line maps (not presented here) show any distinct features at the same
locations of the inner bars. The stellar velocity dispersion map for NGC\,5850
was previously shown by \citet{2004A&A...421..433M}, but no hollows are seen
in there; we believe this is because the smaller FoV of the MPFS instrument used
in those obervations.\looseness-2

\section{Discussion: the origin of the $\sigma$-hollows}
Since the $\sigma$-hollows of the four galaxies are observed exactly at the ends of the
inner bars, it seems that they are related to the inner bar itself and not to any
other structural or kinematical component. The aim of this section is to investigate the origin
of the $\sigma$-hollows, so in the following we discuss some possible explanations for these
observations.

\subsection{The presence of an inner disk}
Stellar inner disks could, in principle, explain a decrease in velocity dispersion in
the central regions of galaxies, as happens with the $\sigma$-drops seen in
\citealt{2001A&A...368...52E}. However, the observed alignment of the
$\sigma$-hollows with the major axis of the inner bar cannot be accounted for
with the presence of an inner disk whose major axis is usually well aligned with the
main photometric major axis of the galaxy. Moreover, the decrease in $\sigma$ in our
galaxies does not occur at the same locations as the $\sigma$-drops seen by
\citet{2001A&A...368...52E}; in fact we find that the velocity dispersion reaches a
maximum value at the center. The possibility of a gaseous inner disk is
discarded because the effect we see in our data is purely stellar.

\subsection{A young stellar population component}
A young stellar component that has acquired the kinematics of the cold
gas it was formed from could also explain the decreases in the velocity dispersion values.
This young population might not necessarily be associated with a different
structural component (e.g. a nuclear star forming ring). 
We tested this hypothesis by performing 
a preliminary stellar population analysis of the different structures to investigate 
the presence of young stars (an extensive stellar population analysis will be published 
elsewhere). However, we do not find any evidence of the presence of a particularly young 
stellar population at the $\sigma$-hollows locations.

\subsection{A matter of contrast}
We now focus on the immediate surroundings of the inner bar. In the central
parts of these galaxies we have the combination of a component with typically
high velocity dispersion (i.e. a bulge) and the inner bar with its ordered
motion and thus a low $\sigma$. We propose that the presence of the
$\sigma$-hollows is due to the contrast between the velocity dispersion of these
two components. Since the velocity dispersion profile and the luminosity profile
of the bulge decreases outwards, we can only see this effect in the outer parts
of the inner bar, where the bulge is not totally dominating the flux. Therefore
the amplitude of the hollows will depend on the relative contribution of the
bulge to the total luminosity at the extremes of the bar and on the difference
of velocity dispersions between the bulge and the inner bar. Irrespective of the
relative luminosity of the two components at the edges of the inner bar, if
there were no differences between the velocity dispersions we would not be able to
find any $\sigma$-hollows. Assuming that there is a significant difference in the
$\sigma$ values at these points, a very extended bulge may have the inner bar
embedded in it, so we would expect very deep $\sigma$-hollows in this case. On
the contrary, if the bulge were less extended these $\sigma$-hollows would not
show up.

To test this idea we used the code FTM 4.4 (updated  version) from
\citet{1994ApJ...424...84H} to perform N-body self-consistent three-dimensional
numerical simulations starting with a (classical-like) bulge and an exponential
disk. First, we relax the bulge, in order to have a ``quiescent''  start, and then we
introduce an exponential self-graviting rotating disk. The bulge proceeds to 
evolve almost steadily, contrary to the disk, which develops a bar instability. 
Here, we use simulations with a single bar to mimic the main components in the region of
interest: the central kiloparsec. This simple approximation is enough to
reproduce the $\sigma$-hollows: they appear at the ends of the bar for the cases
in which the contrast is sufficient. These $\sigma$-hollows last as long as the
bar stays with a low $\sigma$. Stellar bars in general could be
heated-up by the buckling instability, which is milder in the presence of gas
\citep{1998MNRAS.300...49B,2007ApJ...666..189B} and in weak bars \citep{2004ApJ...613L..29M}.
Since the four observed inner bars presented here are not very strong 
\citep{2004A&A...415..941E} and there is some gas in these regions, most likely
they will not buckle. Therefore, we expect the $\sigma$-hollows to be long-lived structures.

\begin{figure}[ttt]
\begin{center}
\includegraphics[width=8cm,angle=0]{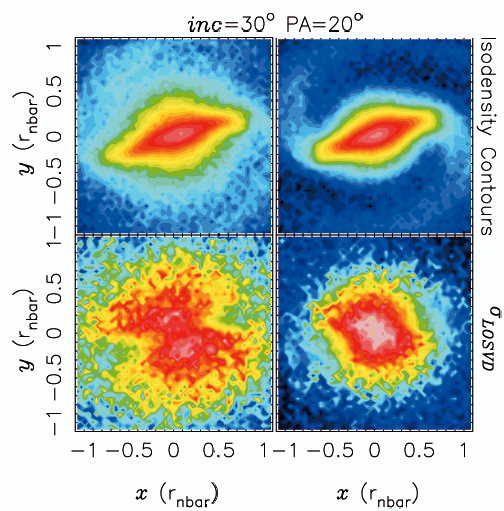}
\caption{Numerical simulations including a bulge, a bar and a disk. Top panels: 
iso-density contours; lower panels: LOSVD. On the left is the case for a bulge 
big enough to match the size of the bar: the corresponding LOSVD map 
shows clearly the $\sigma$-hollows. On the right is the case of a very small
bulge, whose luminosity drops quickly with radius such that its contribution is not 
relevant at the ends of the bar. 
The $\sigma$-hollows in this case become difficult to detect.}
\label{fig:model}
\end{center}
\end{figure}

In Figure~\ref{fig:model} we illustrate two of our models: one case of a bulge
large enough to match the length of the bar, where we are able to reproduce the
$\sigma$-hollows, and one case with a smaller, less extended bulge where the
velocity dispersion profile is almost unchanged. The simulations shown here use
a galaxy inclination of 30$^{\circ}$ (as a representative case of our 
observations) and a position angle of 20$^{\circ}$ for
the bar. However, we repeated the tests with different observational conditions,
getting similar results in all cases. Note however that recent numerical
simulations of double-barred systems by \citet{2007arXiv0711.0966S} do not show
any evidence of the presence of the $\sigma$-hollows in their velocity
dispersion maps. A possible explanation for this absence is the fact that in
those simulations there is not sufficient velocity dispersion difference between
the resulting structural components.

\section{Conclusions}
We have presented 2D kinematical maps of double-barred early-type galaxies
covering the whole nuclear region. We have found $\sigma$-hollows appearing at
the ends of the inner bar for our four objects. The main result presented in
this paper is that these hollows are signatures of inner bars, which
indicates that it they must have significantly lower velocity dispersions than the
surrounding bulges. This means that inner bars are cold systems, and are
not related to triaxial bulges \citep{2004ARA&A..42..603K}. Moreover,
the observations presented here put constraints on the degree of rotational
over pressure support for these structural components. Finally, since
the $\sigma$-hollows are not due to other structural or kinematical components
(e.g. inner disks), they may be used to identify inner bars from a purely
stellar kinematic analysis.\looseness-2

It is likely that previous numerical simulations were unable to predict the
observed $\sigma$-hollows because they did not include a sufficiently
dynamically hot component (i.e. classical bulge). However, we have been able to
reproduce this feature with simulations including a hot bulge, a single small
bar and a disk. The $\sigma$-hollows can be seen due to a contrast between the
high velocity dispersion of the bulge and the low velocity dispersion of the
bar. The amplitude of the hollows depends on the relative contributions to the
total flux and the relative sizes of the bulge and the inner bar. While
young stellar populations can help to lower the stellar velocity dispersion, no
clear signatures of them are found in our sample. Based on the pieces of
evidence presented in this paper, we think that the $\sigma$-hollows 
are simply a matter of contrast.
\\

\acknowledgements The authors thank E.\,Emsellem, R.\,McDermid and R.\,Peletier for
their help concerning data reduction and analysis. Their comprehensive comments,
together with those from J.\,A.\,L.\,Aguerri, P.\,T.\,de\,Zeeuw, J.\,Knapen and I.\,Shlosman,
have greatly helped to improve earlier drafts of the manuscript. The authors 
are indebted to M.\,Beasley for his help correcting the English, to
the \sauron\ team for their support for the realisation of 
this project and to the anonymous referee for constructive comments. 
AdLC gratefully acknowledges partial financial support from
ESA/ESTEC and a European Union EARA Fellowship for visiting the Leiden
Observatory. This work has been supported by the grants AYA2004-03059 and
AYA2007-67752-C03-01, funded by the Spanish Ministry of Education and Science.


\clearpage

\twocolumn

\end{document}